\documentclass[runningheads]{llncs}

\usepackage{graphicx}
\usepackage{multirow}
\usepackage{xcolor}
\usepackage{amssymb}
\usepackage{amsmath}
\usepackage{ulem}
\usepackage[hidelinks]{hyperref}
\usepackage{lscape}
\usepackage{adjustbox}
\usepackage{rotating}
\usepackage{enumitem}

\definecolor{dkgreen}{rgb}{0,0.6,0}
\definecolor{gray}{rgb}{0.5,0.5,0.5}
\definecolor{mauve}{rgb}{0.58,0,0.82}

\usepackage{listings}
\usepackage{multicol}
\lstdefinelanguage{xes}{%
  language     = XML,
  morekeywords = {trace, int, key, value, event, date, string, float, boolean},
   morecomment = [s]{/*}{*/},
}

\newif\ifcomment \commentfalse								% use this to comment out unwanted latex code

\graphicspath{{./figs/}}

\begin{document}
\title{
%Discovering and Analyzing Work Profiles of \\ Resource Groups using Event Logs
%Mining Resource Group Behavior \\ from Event Logs
%Mining Work Profiles of Resource Groups \\ using Event Logs
%Mining Work Profiles of Resource Groups\\ for Workforce Analytics}
AMORETTO: A Method for Deriving IoT-enriched Event Logs
}

%\ifcomment
\author{
    Jia~Wei\inst{} \and
    Chun~Ouyang\inst{} \and
    Arthur~H.M.~ter~Hofstede\inst{} \and
    Catarina~Moreira\inst{}
}
\authorrunning{J.~Wei et al.}
\institute{
    Queensland University of Technology, Brisbane, Australia 
}
%\fi
%

\maketitle % typeset the header of the contribution
\begin{abstract}
\vspace*{-1.25\baselineskip}
Process analytics aims to gain insights into the behaviour and performance of business processes through the analysis of event logs, which record the execution of processes. 
%Event logs are multi-dimensional time sequence data that record the execution of processes. 
With the widespread use of the Internet of Things (IoT), IoT data has become readily available and can provide valuable context information about business processes. As such, process analytics can benefit from incorporating IoT data into event logs to support more comprehensive, context-aware analyses. However, most existing studies focus on enhancing business process models with IoT data, whereas little attention has been paid to incorporating IoT data into event logs for process analytics. %so far 
%main summary of this work contribution
Hence, this paper aims to systematically integrate IoT data into event logs to support context-aware process analytics. To this end, we propose AMORETTO – a method for deriving IoT-enriched event logs. Firstly, we provide a classification of context data, referred to as the IoT-Pro context classification, which encompasses two context dimensions: IoT context and process context. Next, we present a method for integrating IoT data with event logs, guided by IoT-Pro, to yield IoT-enriched event logs. To demonstrate the applicability of AMORETTO, we applied it to a real-life use case and examined whether the derived IoT-enriched event log sufficed to address certain specific analytical questions. 
\vspace*{-.75\baselineskip}

\end{abstract}

% Mining event logs of business process execution provides a promising way towards addressing the above challenge. 
%, as end-to-end processes connect the performance of different employee groups with the business outcomes of an organization. 
%
%
\section{Introduction} \label{sec:intro}
\vspace*{-0.5\baselineskip}
%Background
%% process analytics
% As organisations have expanded their use of information systems, business process data has become more accessible.
Process analytics, which analyses process data to obtain insight into the behaviour and performance of business processes, has been gaining popularity in practice in recent years.
%% process mining
In fact, process mining has become a popular technique for conducting business process analytics over the last decade~\cite{erdougan2016process} and is dedicated to discovering, monitoring and improving processes. 
% The main idea of process mining is to extract knowledge about business processes based on process execution data recorded in event logs~\cite{vanderAalst2016ProcessMD}. 
%% event log & context
Event log data forms the foundation of process mining and captures multidimensional time-series data that characterises the execution of processes. Among others, event logs contain context associated with business processes, which has been exploited by numerous studies for the purpose of process analytics.

%% IoT & context
Over the last few decades, the concept of the Internet of Things (IoT) has gained a lot of attention in both academia and industry. It enables the interconnection and communication between everyday objects via the Internet~\cite{dorsemaine2015internet}. With the widely increased application of the IoT, more and more data has become readily available. IoT data can capture various elements of context related to business processes. For example, it can capture up-to-date information about physical objects and human resources associated with process execution and the location and environment where processes are executed~\cite{PereraZCG14}.

%Motivation
Process analytics based on event logs can benefit from integration with IoT data to provide a more comprehensive, context-aware analysis. However, most existing research has focused on how IoT data can be used to enhance business process models. Thus far, the use of IoT data to improve process analytics has received little consideration. Recently, a manifesto describing the mutual benefits and challenges regarding the interaction between the IoT and business process management was published~\cite{Janiesch2017TheIM}, %. In this manifesto, 
and the authors identified 16 challenges with this interaction. In this work, we aim to address one of these challenges: \textit{bridging the gap between IoT data and event logs for process mining}. Since IoT data is normally low-level data, it cannot be directly integrated into event logs as these contain relatively high-level information about process executions~\cite{ValderasTS22}. This leads to the following research question for leveraging the IoT for process analytics: \textit{How to systematically integrate IoT data with event logs in order to obtain IoT-enriched event logs to further support context-aware process analytics?}

%Scope

%Contribution
In this paper, we propose AMORETTO, \textit{A Method fOR dEriving~ioT-enriched evenT lOgs}. 
Our first step is to provide a classification of context data referred to as the \textit{IoT-Pro} context classification, which includes two dimensions of context: IoT and process. Secondly, we present a method for integrating IoT data with event logs, guided by IoT-Pro. 
Through AMORETTO, process analysts can gain a comprehensive understanding of the process in which IoT devices are applied, thereby enhancing process analysis capabilities. 
To demonstrate the applicability of AMORETTO, we apply it to a real-world use case, generate IoT-enriched event logs driven by analysis scenarios, and examine whether the obtained event logs contain sufficient information for answering specific analytical questions.
%the IoT-Pro context classification. 
%analyse certain scenarios to 

% Not sure the following is needed
% \textcolor{red}{Our work contributes to ....}

%% -- REMOVE TO SAVE SPACE FOR SUBMISSION -- %%
%The rest of the paper is structured as follows. Sect.~\ref{sec:background} introduces the terminology used in this study and reviews related work. Sect.~\ref{sec:method} presents the proposed method --- AMORETTO, and Sect.~\ref{sec:application} discusses its application to a real-life use case. Sect.~\ref{sec:conclusion} concludes the paper and provides suggestions for future work.		% 1-2 pages
\section{Background and Related Work}
\label{sec:background}

In this section, we set the scene for the remainder of the paper through the introduction of concepts and terminology, the presentation of an illustrative example, and a discussion of related work.

\vspace*{-0.5\baselineskip}
\subsection{Concepts and Terminology}

A \textit{process} consists of a series of activities and decisions, involving people and objects, with a specific goal or outcome. An \textit{event log} captures information about the execution of process instances (\textit{a.k.a.} \textit{cases}). % a collection of cases (that is, process instances) in the process. 
A case is uniquely identified and can be viewed as a sequence of \textit{events}. where each event is described by a set of \textit{attributes}, such as case id, activity name, timestamp, etc.~\cite{Aalst16}. Table~\ref{tab:table_eventlog} shows an example of an event log. Each row represents an event capturing the execution of an activity that occurred during a patient’s visit at an emergency department (ED), and each column represents an attribute of the event.

There are \textit{static} attributes, which carry the same value within the same case, and \textit{dynamic} attributes of which the value can change from event to event~\cite{TeinemaaDRM19}. These attributes can carry \textit{context} relevant to a business process, e.g., a patient's acuity level assessed at triage during an ED visit. With the use of the IoT on business processes, IoT data contributes a richer context to process execution. For example, a patient's temperature and heart rate can be measured more frequently and precisely from the associated body sensors. 

%In addition, an event log may contain process contexts that are represented by attributes that have an impact on process execution. It may also include IoT contexts that are captured by IoT data that influence process execution. In this work, we refer to the IoT-Pro context as the integration of the process and IoT contexts. Guided by the IoT-Pro context classification, we aim to generate an IoT-enriched event log which refers to an event log enriched by information captured by IoT data.
% and relevant for process execution?

%\subsection{Illustrative Example}

\begin{table}[t!!!]
\resizebox{\textwidth}{!}{%
\begin{tabular}{|l|l|l|l|l|l|l|l|}
\hline
{\textbf{case id}} & \multicolumn{1}{c|}{{\textbf{timestamp}}} & \multicolumn{1}{c|}{{\textbf{activity}}} & \multicolumn{1}{c|}{{\textbf{temperature}}} & \multicolumn{1}{c|}{{\textbf{heartrate}}} & \multicolumn{1}{c|}{{\textbf{pain}}} & \multicolumn{1}{c|}{\textbf{acuity}} & \multicolumn{1}{c|}{\textbf{chiefcomplaint}} \\ \hline
0001  & 29/03/2110 18:36  & Enter the ED &   &    &    &    &   \\
0001  & 29/03/2110 18:36  & Triage in the ED & 97 & 68 & 5 & 3 & R Inguinal pain \\
0002  & 29/03/2110 19:37  & Enter the ED &   &    &    &    &   \\
0002  & 29/03/2110 19:37  & Triage in the ED & 99.8 & 110 & 0 & 2 & ETOH  \\
0002  & 29/03/2110 19:38  & Vital sign check & 99.8 & 110 & 0 &   &    \\
0001  & 29/03/2110 20:29  & Medicine reconciliation &    &   &   &   &   \\
0002  & 30/03/2110 06:58  & Discharge from the ED  &    &   &   &   & \\
0001  & 30/03/2110 10:21  & Vital sign check & 97.9  & 60  & 2  &   & \\
0001  & 30/03/2110 11:56  & Discharge from the ED &   &    &   &   & \\
0003  & 30/03/2110 19:40  & Vital sign check  & 98.8  & 80 & 0  &   & \\
0003  & 30/03/2110 19:40  & Enter the ED &  &  &   &   &  \\
0003  & 30/03/2110 19:40  & Triage in the ED  & 98.8  & 80 & 0  & 4  & EXPOSURE  \\ \hline
\end{tabular}
}
\vspace{.05\baselineskip}
\caption{A snippet of an event log capturing the process of a patient's ED visit~\cite{mimicel}.}
\vspace*{-2.15\baselineskip}
\label{tab:table_eventlog}
\end{table}

%depicts a snippet of an event log capturing a business process in the healthcare domain. Specifically, this process describes the journey of a patient at an emergency department (ED)~\cite{mimicel}. 
%This event log contains process contexts that can be used to analyse the performance of the ED process (i.e., length of stay), e.g. the patient’s self-reported pain level, their acuity level, and their chief complaint. In addition, it also includes IoT contexts such as temperature and heart rate that were collected by body sensors connected to the patients. 
%For example, an event could be a patient entering the emergency room at a certain time during a specific stay. 
% The process starts with the arrival of the patient and on entry to the emergency department the patient is triaged and depending on the severity of the patient, the patient is given vital signs, or asked about current medication history, or given medication directly, or is given some or all of the above activities at the same time. After this, the patient may be discharged from the ED or transferred to another care unit or admitted to hospital.
%
%Overall, this event log captures information about a collection of cases that represent different ED visits. 

\vspace*{-.5\baselineskip}
\subsection{Related Work}

% one paragraph for process context
% one paragraph for IoT context
% integration of IoT-BPM (combine two domains)
% Business processes that describe a structured flow of activities~\cite{RosemannRF08} in an organisation used to be designed in a static and prescribed manner and isolated from the environment~\cite{ploesser2009learning}. They were not flexible enough to capture the changes that occurred in the environment in which the process was embedded, which led to poor process performance and further a lack of certain analytical capabilities of the process. 

\vspace*{-.25\baselineskip}
\subsubsection{Business process context} 
There has been an increasing interest in taking into account relevant context for process analytics. Contextual variables can capture changes in different environmental conditions and the specific case properties, %handled by the process, 
which may affect the execution of a business process~\cite{ploesser2009learning}. Researchers have presented various views on the notion of business process context. In~\cite{van2012process} van der Aalst and Dustdar propose a classification of different contextual factors to consider for process mining. 
%that may influence the execution of a business process. %in process mining. 
Rosemann et al.~\cite{RosemannRF08} present different layers of context from the perspective of business operations related to processes. 
%, namely Immediate, Internal, External and Environmental, 
Brunk~\cite{Brunk20} proposes a taxonomy that discusses dimensions of business process context and characteristics for predictive process analysis, covering both classifications mentioned above in a broader scope beyond business processes.  
%: time, structure, origin, relevance, process relation, and runtime behaviour. 
%%% [This may be moved to Section 3] 
%Compared to the context classification proposed by~\cite{van2012process}, this model captures a greater number of contexts outside the process, while contexts relating to the core process are only placed in the central layer. 
% Unlike above taxonomies that focus on conceptualisation and classification of business process contexts~\cite{van2012process,RosemannRF08}, 

\vspace*{-.75\baselineskip}
\subsubsection{IoT context}
Studies on classification of context carried by IoT data (referred to as \textit{IoT context} for short) focus on conceptual and/or \mbox{operational perspectives~\cite{BunningenFA05}.} 
From a conceptual perspective, Schilit et al.~\cite{schilit1994context} identify three essential aspects of IoT context: location-related, resource-related and people-related information. 
Abowd et al.~\cite{AbowdDBDSS99} propose four main types of IoT context: location, identity,~time, and activity. 
From an operational perspective, the emphasis is on the varying data granularity. %of IoT context. 
For instance, Abowd et al.~\cite{AbowdDBDSS99} propose to classify IoT context as primary and secondary. Henricksen et al.~\cite{henricksen2002modeling} categorise IoT context based on their persistence and data sources: sensed, static, profiled and derived. 
Based on a survey of existing studies, Perera et al.~\cite{PereraZCG14} present a refined classification of IoT context from both conceptual and operational perspectives. 
However, there are still missing categories of IoT context, such as environment-related information (e.g., weather, humidity, atmospheric pressure), in the existing classifications.

\vspace*{-.75\baselineskip}
\subsubsection{IoT and process analytics}
Recently, the integration of the IoT and business process management (BPM) has gained in attention. 
Stoiber and Sch{\"{o}}nig~\cite{StoiberS22} introduce a taxonomy of characteristics of industrial IoT applications for business process improvement. 
Valderas et al.~\cite{ValderasTS22} apply Business Process Model and Notation (BPMN) and microservices to model and execute IoT-enhanced business processes. 
Song et al.~\cite{SongCVHW22} present an approach to leverage IoT data for improved decision making in a business process. 
Where attempts have been made to integrate the IoT with BPM, most of the existing work focused on how to enhance business process design in the presence of the IoT, e.g., to extend process models with IoT activities and data. 
Few studies explore the use of IoT data for process discovery. For instance, Koschmider et al.~\cite{KoschmiderJM20} investigate how to discover sensor-level processes from sensor events and raw IoT data. 
%There are also some studies that use IoT data to improve process discovery and conformance checking, however, most of these studies suggest discovering processes directly from IoT data, i.e., discovering sensor-level processes.  
%
%The combination of IoT and BPM has been examined in a number of studies. 
%explore the impact of implementing IoT applications on business process improvement (BPI). 
%Some studies focus on sub-areas of business process management (BPM) and process mining. For instance, 

%% FIND A SUITABLE PLACE TO ADD THIS ... %%
%IoT can enrich BPM through continuous data sensing and physical actuation, and with up-to-date IoT data, a more holistic view of the process can be obtained, which further enables enhanced process analytics~\cite{Janiesch2017TheIM}. %enhancing process analysis capabilities %However, 

Integrating the IoT and BPM is a relatively new topic and poses many open challenges. 
In a recent manifesto, Janiesch et al.~\cite{Janiesch2017TheIM} identified 16 challenges arising from interactions between the IoT and BPM. In this work, we %specifically 
aim to~address one of these challenges --- to bridge the gap between sensor data and event logs~\cite{Janiesch2017TheIM}. 
In this regard, the most relevant work is Bertrand et al.~\cite{BertrandWS21}, which applies an event-driven approach to address the gap. The authors propose three types of events and their relations, with an aim to establish a link between the IoT and process mining. More specifically, an `IoT event' captures a change in the status of a physical object or environment monitored by an IoT device (e.g., a sensor), a `context event' is derived from an IoT event that influences the execution of a process, and a `process event' is derived from an IoT event that indicates a state change in the transactional lifecycle of an activity in a process. Different from their work~\cite{BertrandWS21}, our research focuses on (i) how to classify IoT context in relation to different levels of process context (named \textit{IoT-Pro context}) and (ii) how to integrate IoT context with event logs.  

\section{Method for Deriving IoT-enriched Event Logs}
\label{sec:method}

This section describes AMORETTO in detail. Firstly, we discuss how the IoT-Pro context classification was derived. Secondly, we illustrate how to integrate IoT data into event logs using IoT-Pro as guidance.

\subsection{Classifying IoT-Pro Context}
\begin{figure}[t!!!]
%    \vspace*{-1.2\baselineskip}
    \centering
    \includegraphics[width=.8\textwidth]{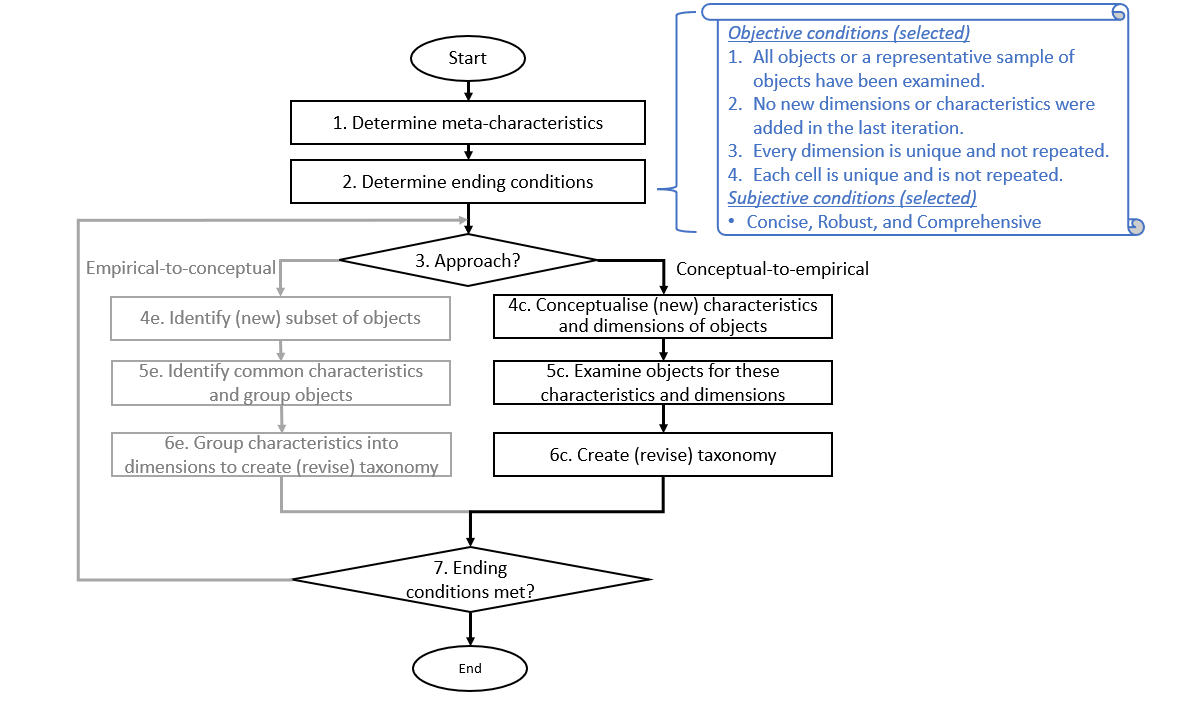}
    \vspace*{-\baselineskip}
    \caption{The taxonomy development method introduced by Nickerson et al.~\cite{Nickerson2013AMF}}
\label{fig:taxonomy_method}
\vspace*{-\baselineskip}
\end{figure}

% \subsubsection{Research Method} 
%In this work, 
We adopt the taxonomy development method introduced in~\cite{Nickerson2013AMF}, as depicted in Fig.~\ref{fig:taxonomy_method}, to derive the IoT-Pro context classification. 
%It provides systematic and rigorous guidance for the design process of taxonomy development. In particular, the method includes a number of iterative steps, taking into account both theoretical studies and empirical data, with the aim of creating a comprehensive taxonomy. 
%% --- Describe the method itself
First of all, it is required to determine the characteristics of the objects of interest based on the purpose of the taxonomy (Step~1) and the conditions for terminating the iterations of the development steps (Step~2). Next, the method proposes two paths towards the development of a taxonomy (Step~3). One can choose either the ``conceptual-to-empirical'' (Steps~4c, 5c, 6c) or ``empirical-to-conceptual'' (Steps~4e, 5e. 6e) to start the development, depending on the depth of knowledge about the domain and empirical data available. When there are few experimental data available yet a great deal of relevant research in this field, it is proper to begin with a ``conceptual-to-empirical'' approach. Alternatively, when the researcher is unfamiliar with the area but has access to a considerable amount of empirical data, an ``empirical-to-conceptual'' approach is recommended. These will then be iterated until all of the ending conditions have been satisfied (Step~7).

% 
% In specific, we select the following objective ending conditions: ``All objects or a representative sample of objects have been examined'', ``No new dimensions or characteristics were added in the last iteration'', ``Every dimension is unique and not repeated'' and ``Each cell is unique and is not repeated''. The subjective ending conditions define that the taxonomy should be: \textit{concise}, \textit{robust} and \textit{comprehensive}.
%
In our work, the IoT-Pro context classification aims to provide users with guidance on incorporating relevant and meaningful IoT data into event logs (Step~1). 
We then employ three subjective and four objective conditions from~\cite{Nickerson2013AMF} as ending conditions (Step~2). These can be considered as the \textit{ex-ante} evaluation criteria for assessing the proposed classification. 
Next, we mainly adopt the ``conceptual-to-empirical'' approach for deriving an IoT-Pro context classification (Steps~4-6c). Whilst there already exist several classifications of process context and IoT context, respectively, experimental data of event logs with relevant IoT context are scarce. 
As a result, we review and refine the selected existing context classifications and synthesise them to the IoT-Pro context classification.
%While conceptual knowledge of such integration is sorely lacking, process context and IoT context have been extensively studied. 
% Thus, we choose to refine  existing context classifications on business process contexts introduced in~\cite{van2012process} because it specifically discusses contexts that impact the process execution but were neglected in process mining analysis.
%Therefore, we review and refine existing classifications in the business process context and IoT context. 
% With the IoT's contribution Big Data, more data has emerged that could not be included at the current process context levels. Sensor data, for example, is simply classified into the external context in existing process context classifications. In this work, the classification of IoT settings is approached from a conceptual standpoint. Contexts of various granularities can be obtained implicitly by context derivation, such as preprocessing or aggregation. At this stage, we create an initial taxonomy that incorporates both the process and IoT context.

\vspace*{-.75\baselineskip}
\subsubsection{IoT-Pro context classification}
encompasses two dimensions where one captures a refined proposal of process context levels based on~\cite{van2012process} and the other specifies a classification of IoT context adapted from~\cite{PereraZCG14}.

% process context dimension
In~\cite{van2012process}, van Der Aalst and Dustdar propose four levels of process context. 
\textit{Instance context} captures information about the properties of cases that impact process execution, e.g., the acuity level of a patient during an ED visit (Table~\ref{tab:table_eventlog}). 
\textit{Process context} considers the properties associated with a process, regardless of any specific cases, e.g., the number of patients waiting in the triage. % Rather than looking at one process instance in isolation, it focuses on multiple instances within the process. It provides an aggregated view of the context at the instance level. 
\textit{Social context} captures information about the entities related to the process in the organisation, e.g., the interaction between nurses and doctors.
\textit{External context} includes factors outside the control of the organisation, e.g., ED medical procedures.

There are a few issues to address in the above proposal. 
\textit{First}, the level of instance context is too coarse, as it contains both context specific to cases (e.g., those carried by static attributes) and context specific to process events (e.g., those carried by dynamic attributes). Therefore, we decide to subdivide this level into two levels: \textbf{instance context} which contains context specific to cases only, and \textbf{event context} specific to process events only. 
\textit{Second}, with the application of the IoT on business processes, it is necessary to capture the context carried by raw IoT data, i.e., data collected by IoT sensors, relevant to process execution. Hence, we introduce a new level called \textbf{sensor context}. 
\textit{Third}, the name `social context' reflects an emphasis on capturing how people work together in a process. As BPM research has evolved, the term `\textbf{organisational context}' has become more appropriate. 
\textit{Fourth}, the external context contains both environment-related context, such as weather conditions which can be captured by IoT sensors, and legislation that needs to be deployed in an organisation. To this end, we disassemble the level of external context by relocating the relevant information into the sensor context and organisational context accordingly.

In~\cite{PereraZCG14}, Perera et al. present a classification of four categories of IoT context: \textit{location}, \textit{identity}, \textit{time} and \textit{sensor activity}. 
This classification raises a \mbox{few issues.} 
First, as our study concerns the context carried by IoT data rather than how sensors operate, the context of `sensor activity' is beyond the scope of this work. 
Second, we introduce two missing types of IoT context, namely `physical object' and `environment'. The proposal of physical objects is in line with the definition of IoT~\cite{Janiesch2017TheIM}, which ``\textit{is the inter-networking of physical objects (the things), such as embedded systems with electronics hardware, software, sensors, actuators, and network connectivity}''. 
The `environment' context specifies the surroundings or conditions captured by sensors. 
Below, we propose a classification of IoT context.

%When considering the application of IoT on business processes, 
% \textit{Location} describes where sensing takes place, of which information can be captured by, e.g., GPS coordinates. 
% \textit{Identity} specifies who is involved, of which information can be captured by, e.g., RFID tags. 
% \textit{Time} specifies when sensing happens, which can be captured by, for example, a timer. 
% \textit{Activity} describes what sensing activity is happening, which can be captured by, for example, a physical object sensor.
%
\begin{itemize}[topsep=0pt]
    \item \textbf{Physical Object}: \textit{what} object is involved in the execution of a process? E.g., medicine information captured by medication barcode.
    %It captures information related to the physical objects associated with process execution, 
    \item \textbf{Location}: \textit{where} is the associated process activity taking place? E.g. geographical information captured by GPS coordinates.
    %It captures geographical information about the process. e.g., GPS coordinates.
    \item \textbf{Time}: \textit{when} is the associated process activity happening? E.g. real-time temporal information captured by a timer.
    % It captures real-time temporal information, e.g., a timer
    \item \textbf{Identity}: \textit{who} is involved in the process execution? E.g., a patient's information captured by a barcode placed on the wristband.
    % It captures information about the identity of those involved in the process, e.g., RFID tags
    \item \textbf{Environment}: this process is \textit{influenced by} what environment? E.g., weather conditions captured by temperature and humidity sensors. 
    
    % It captures information about the elements of the environment in which the process takes place.
\end{itemize}

% need to provide the table which captures the interaction between two dimensions.
Finally, we propose the IoT-Pro context classification through the synthesis of the above two dimensions of context. Table~\ref{tab:refined_taxonomy} provides a detailed definition of IoT-Pro where each cell specifies the interaction between a process context level and an IoT context category.

%captured by each cell (a combination of contexts in two dimensions) is provided in .

\begin{table}[h!]
%\vspace*{-\baselineskip}
\begin{adjustbox}{max width=\textwidth}
\begin{tabular}{|c|p{3.3cm}|p{3.3cm}|p{3.3cm}|p{3.3cm}|p{3.3cm}|}
\hline
\multicolumn{1}{|l|}{}  & \multicolumn{1}{c|}{\textbf{\begin{tabular}[c]{@{}c@{}}Physical Object\\ (What)\end{tabular}}}                                                                      & \multicolumn{1}{c|}{\textbf{\begin{tabular}[c]{@{}c@{}}Location\\ (Where)\end{tabular}}}                    & \multicolumn{1}{c|}{\textbf{\begin{tabular}[c]{@{}c@{}}Time\\ (When)\end{tabular}}}                       & \multicolumn{1}{c|}{\textbf{\begin{tabular}[c]{@{}c@{}}Identity\\ (Who)\end{tabular}}}                                         & \multicolumn{1}{c|}{\textbf{\begin{tabular}[c]{@{}c@{}}Environment\\ (Influenced by)\end{tabular}}} \\ \hline \hline
\begin{tabular}[c]{@{}c@{}}\textbf{Organisational}\\ (Org.) 
\end{tabular} & Org. rules or guidelines applied to physical objects involved in the process execution & Geographical-related org. rules or guidelines applied during the process execution  & Temporal-related org. rules or guidelines applied during the process execution & Org. rules or guidelines applied to human resources involved in the process execution  & Org. rules or guidelines related to environmental requirements influencing the process execution   \\ \hline
\textbf{Process} & Aggregated object info. across multiple cases & Aggregated location info. across multiple cases & Aggregated temporal info. across multiple cases & Aggregated info. about human resources across multiple cases & Aggregated environment info. across multiple cases \\ \hline
\textbf{Instance}  & Object info. associated with a case  & Geographical pattern (e.g., distance, route) associated with a case & Temporal pattern (e.g., seasonal/periodic pattern) associated with a case & Info. about a person (e.g., availability, profile) associated with a case & Environment info. associated with a case \\ \hline
\textbf{Event}  & Object info. associated with one or multiple process events  & Geographical info. associated with one or multiple process events  & Time info. (e.g. point of time, duration) associated with one or multiple process events & Info. about a person who interacts with one or multiple process events & Environment info. associated with one or multiple process events  \\ \hline
\textbf{Sensor}  & Object's sensor data  & GPS coordinates (longitude \& latitude)  & Sensor timing data & Sensor data storing a person's identity & Sensor data capturing environment conditions \\ \hline
\end{tabular}
\end{adjustbox}
\vspace*{.05\baselineskip}
\caption{The proposed IoT-Pro context classification}
\vspace*{-2.15\baselineskip}
\label{tab:refined_taxonomy}
\end{table}

%\subsection{Generating IoT-enriched Event Logs}

%\vspace*{-2\baselineskip}
\subsection{Integrating IoT Data with Event Logs}

\begin{figure}[b!]
 \vspace*{-.5\baselineskip}
    \centering
    \includegraphics[width=0.9\textwidth]{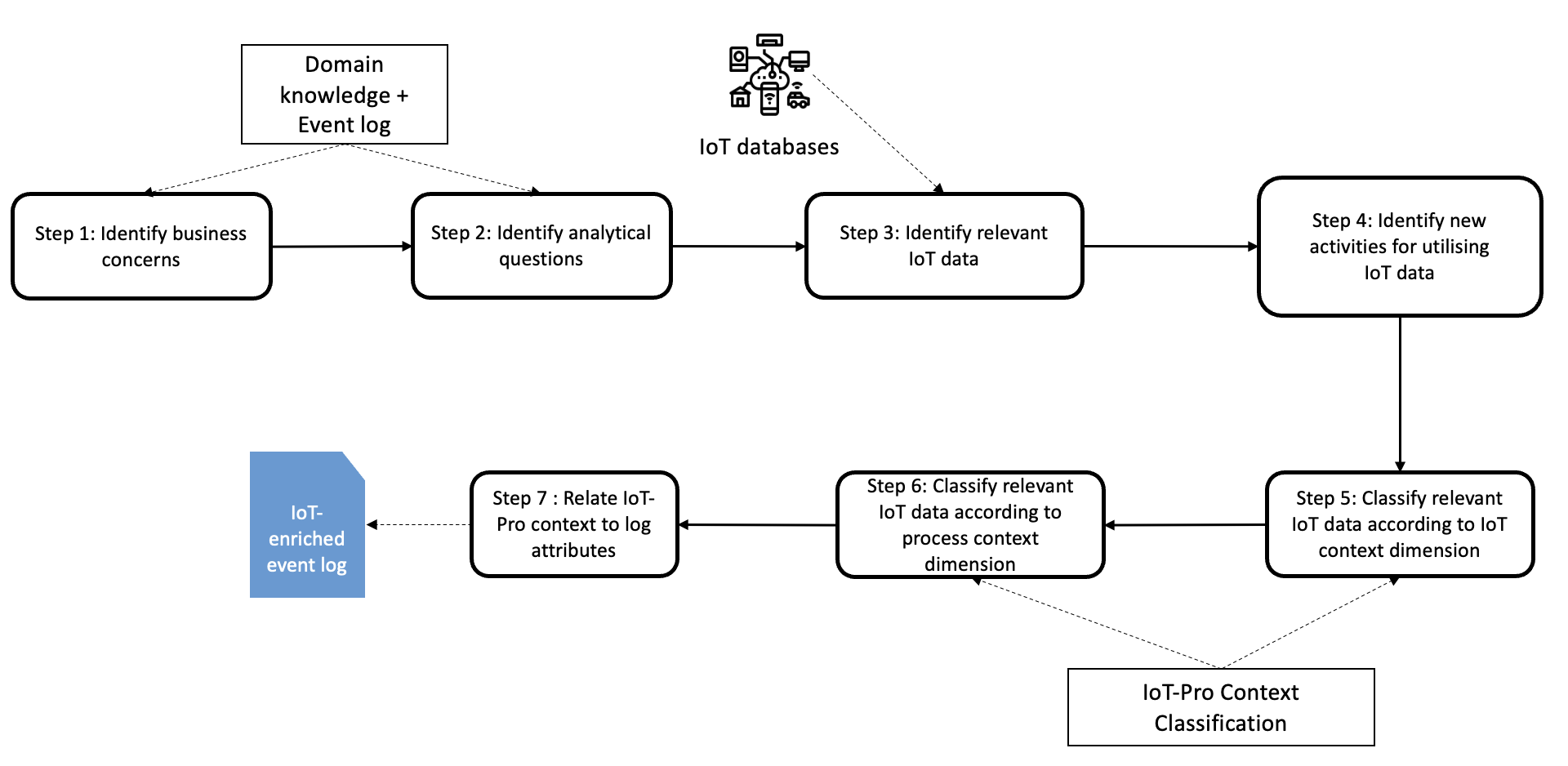}
    \vspace*{-\baselineskip}
    \caption{Overview of the method for generating IoT-enriched event logs}
\vspace*{-\baselineskip}
\label{fig:method}
\end{figure}

We present a method for integrating IoT data with event logs to yield IoT-enriched event logs. The proposed method is an adaption of an existing event log generation guideline for constructing an event log from relational databases~\cite{JansSJ19}. %, which provides systematic procedures for constructing an event log from relational databases. 
As our main objective is to incorporate IoT data into a given event log, we introduce the following modifications to the original guideline in~\cite{JansSJ19}. 
\textit{Firstly}, we remove the steps of extracting event log components from relational databases. 
\textit{Secondly}, we introduce new steps for identifying and utilising relevant IoT data.
\textit{Thirdly}, we propose steps to classify IoT data using the IoT-Pro context classification as guidance.
\textit{Finally}, we provide recommendations for relating IoT-Pro context to event log attributes following existing event logging guidelines. As a result, the proposed method consists of a total of seven steps as shown in Fig.~\ref{fig:method}.

% It is challenging to integrate IoT data directly into the event log since it is stored in a distributed manner~\cite{PereraZCG14}.  

% 3) Since IoT data is low-level~\cite{Janiesch2017TheIM}, it cannot be linked to processes containing high-level knowledge. Consequently, IoT data is included in the event log component by using the proposed IoT-Pro context classification. 4)

\vspace*{-.5\baselineskip}
\paragraph{\textbf{\textit{Step 1.}} Identify business concerns} 
The purpose of this step is to identify key business objectives~\cite{JansSJ19} that are of concern to the relevant stakeholders, e.g., what is expected to be achieved by applying IoT to their business operations. An objective could be to increase process efficiency or to ensure that process performance conforms to predefined rules~\cite{JansSJ19}. 
Hence, the input data for this step includes both an event log of an existing process and process domain knowledge.
% in addition to an event log documenting the existing process, the input data for this step also includes domain knowledge about the process, such as issues in current processes. %business processes.

\vspace*{-.5\baselineskip}
\paragraph{\textbf{\textit{Step 2.}} Identify analytical questions}
This step aims to identify analytical questions pertinent to business concerns identified in Step~1. In order to ensure that all stakeholders have a common understanding, it is essential to clarify key questions that must be answered to address business concerns~\cite{JansSJ19}. According to the process mining manifesto, the extraction of event logs should be driven by specific analytical questions. 

\vspace*{-.5\baselineskip}
\paragraph{\textbf{\textit{Step 3.}} Identify relevant IoT data}
Sensors generate data in real-time, and as a result, IoT data is typically of high volume, high velocity and high variety~\cite{Janiesch2017TheIM}. 
Also, IoT data is often stored in separate databases for different sensors. 
It is important to identify relevant IoT data for answering analytical questions.

% The data in the existing event logs are insufficient to address certain analytical inquiries. 

% The applications of the Internet of Things offer a lot of data to processes that can be utilised to analyse business processes~\cite{PereraZCG14}. On the basis of IoT data, process analysis, execution, and monitoring can result in a more thorough understanding of processes and process optimisation potential~\cite{Janiesch2017TheIM}. 

\vspace*{-.5\baselineskip}
\paragraph{\textbf{\textit{Step 4.}} Identify new activities for utilising IoT data}
The IoT enables business to manage and, if necessary, change the execution of their business processes~\cite{FerrettiS16}. Specifically, activities in business processes may need to be modified in light of the IoT data being collected. Therefore, this step focuses on identifying new process activities that need to be introduced in order to facilitate the use of IoT data discovered in Step~3.

\vspace*{-.5\baselineskip}
\paragraph{\textbf{\textit{Step 5.}} Classify relevant IoT data according to IoT context dimension}
    
% The main objective of the proposed IoT-Pro context classification is to classify contexts in which IoT data is captured in relation to business processes. In particular, 
This step focuses on classifying context captured by IoT data (identified in Step~3) into corresponding categories of IoT context proposed in IoT-Pro, namely \textit{physical object}, \textit{location}, \textit{time}, \textit{identity} and \textit{environment} context.

\vspace*{-.5\baselineskip}
\paragraph{\textbf{\textit{Step 6.}} Classify relevant IoT data according to process context dimension}
This step involves classifying IoT data based on its relevance to process context levels. 
It starts from identifying sensor context against each of the IoT context categories obtained in Step~5. Afterwards, \textit{event context}, \textit{instance context} and \textit{process context} can be derived from the identified sensor context through data aggregation and derivation. As a result, this yields an instantiation of the IoT-Pro context classification.

\vspace*{-.5\baselineskip}
\paragraph{\textbf{\textit{Step 7.}} Relate IoT-Pro context to attributes}
    
The last step is to enrich an event log with new attributes for carrying the IoT-Pro context obtained in Step~6. Such an attribute is specific to cases, if it holds context at the (process) instance level, otherwise it is specific to events, if it holds context at the (process) event level. Following the event logging guidelines for process mining~\cite{Aalst15}, we recommend that when relating IoT-Pro context to attributes for enriching event logs, event data should be as unprocessed as possible, and aggregation should be performed during analysis. In addition, sensitive or private data should be excluded while retaining meaningful data associations.

%Within the IoT-Pro context classification, there are different granularities of contexts, many of which are derivable. This illustrates an issue with attribute selection, i.e. the trade-off between generality and complexity in the generated enriched event logs. For instance, when derived attributes are included in an event log, queries for generating event logs may become complicated, but the process analysis that follows is simple. Adding derived attributes to an event log, however, compromises some generality, as the produced event log can only be used to solve specific process-analytical problems. 

%\vspace*{-0.5\baselineskip}
   % 4-5 pages including figures
\section{Application} 
\label{sec:application}

To demonstrate the applicability of AMORETTO, we apply it to a real-life use case and examine whether the derived IoT-enriched event log can address certain analytical questions via two analysis scenarios.
Below we provide details regarding the application and evaluation of the proposed method.

\vspace*{-0.5\baselineskip}
\subsection{Use Case} %Description}
\begin{figure} [htbp!!!]
    \centering
    \includegraphics[width = .9\textwidth]{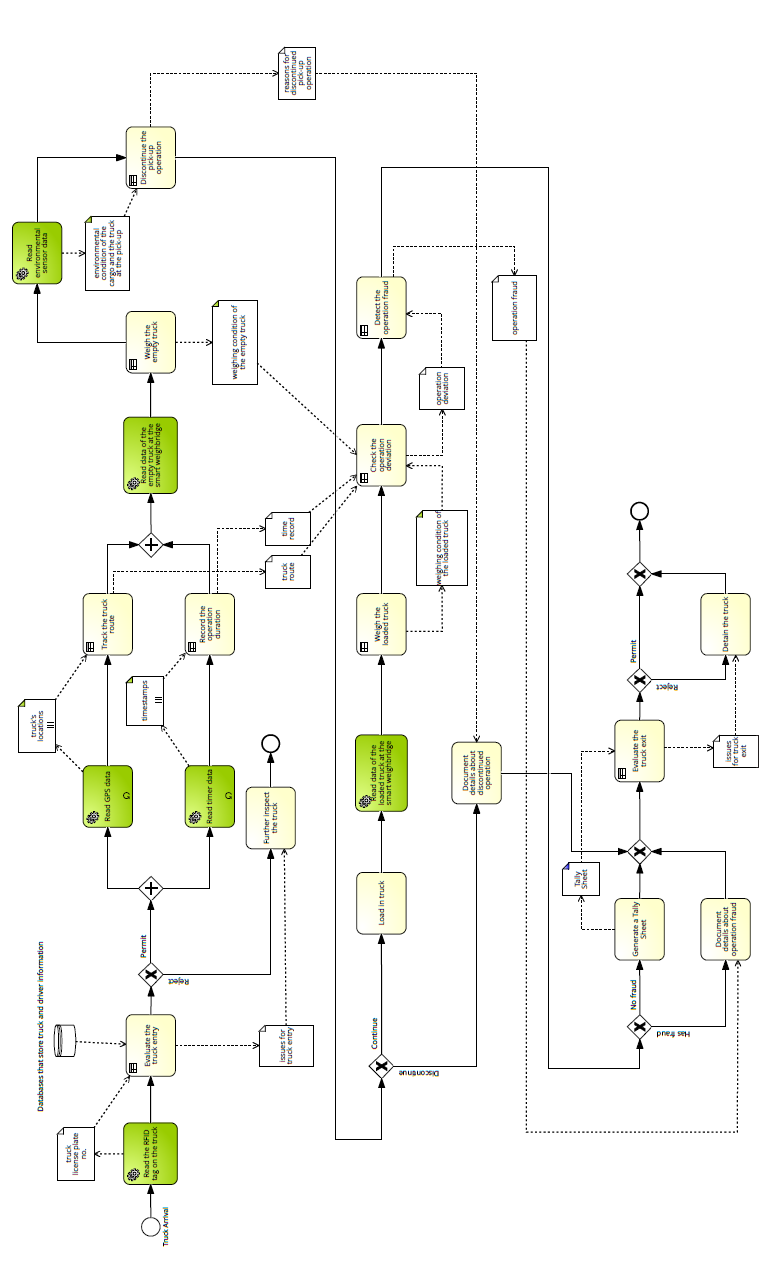}
    \caption{The IoT-enhanced truck pick-up cargo process (a redraw of the truck pick-up cargo process model presented in~\cite{SongCVHW22})}
\label{fig:iot_enhanced_truck_bpmn}
\end{figure}

The real-world use case we explore belongs to the logistics domain and describes the handling of cargo at a bulk port in China~\cite{SongCVHW22}. It specifically describes the process of a truck picking up cargo at the port. The process starts with the arrival of a truck at the port. After that, the truck is verified for eligibility to enter the port. Upon entering the port, the truck is weighed, loaded, and weighed again. A tally sheet is then generated and the truck leaves the port. However, such a pick-up process is frequently plagued by major operational issues, such as loss of cargo and fictitious pick-ups~\cite{Song2020FraudDO}. Although the yard manager suggests manual inspections as a solution to these operational issues, they are costly, subjective, and result in a high rate of delay. Consequently, the port has deployed a number of IoT devices in order to increase process efficiency and to enable better automation of port operations. The collected IoT data provides up-to-date information about the process and relevant context, which in turn impacts decision-making as part of the process.
The process model shown in Fig.~\ref{fig:iot_enhanced_truck_bpmn} describes the IoT-enhanced truck pick-up process. This model is a redraw of the truck pick-up cargo process model presented in~\cite{SongCVHW22}, where green colored sensor activities represent the IoT sensors participating in the process and corresponding data objects represent the input data they provide to the activities of the process (modelled using business rule tasks notations). 
In BPMN~2.0~\cite{object2011business}, service tasks employ particular types of services, such as automated applications. Thus, sensor activities are modelled using the service task notation, given the fact that IoT sensors collect data automatically. 
%Below we demonstrate application of AMORETTO in two scenarios from this use case.

\vspace*{-0.5\baselineskip}
\subsection{Analysis Scenarios}\label{sec:scenarios}

\vspace*{-0.25\baselineskip}
\subsubsection{Scenario 1} is concerned with the issue of fraud in the truck pick-up %cargo
process. Detection of cargo theft has been considered a serious issue at the port, as most cases are not detected in time~\cite{Song2020FraudDO}. 
%Ports and cargo owners rarely detect cargo theft in time, as most cases are detected when there is not enough cargo available for pick-up~\cite{Song2020FraudDO}. 
In addition, bulk cargo theft is difficult to detect manually because port terminals are huge and busy. There are various factors that can lead to uncertainty in the pick-up operation, such as the type of cargo, truck information, and the external environment. Thus, a business concern in this instance (Step~1) could be to improve detection of cargo theft during the pick-up process. A specific analytical question (Step~2) related to this concern could be: \textit{How many incidents of cargo theft have occurred during pick-ups?}

The next step is to determine the IoT data needed to answer this question. According to interviews with experts conducted by Song et al.~\cite{Song2020FraudDO}, most cargo theft occurs from retrofitted trucks. These retrofitted trucks carry heavy filler, which after entering the terminal and weighing the empty trucks, is secretly unloaded at the terminal. The weight deviation of the empty trucks is used to `steal' the cargo. In addition, there appear to be some common features between these thefts, such as the time of operation, the weather, and the locale of the truck licence~\cite{Song2020FraudDO}.
% I do not understand the part about the truck license. How can that play a role?
Therefore, information about the truck (captured by an RFID tag), the location of the truck at the terminal (captured by GPS), the weight of the truck (captured by weight sensors), and the weather conditions that can be captured by rain sensors are all highly relevant IoT data that can be used to detect cargo theft at the port (Step~3).

Step~4 concerns the identification of new activities to facilitate the use of this relevant IoT data. In this scenario, the IoT-enhanced truck pick-up process model has included the activities for using IoT data identified in Step~3. These activities are `check the operation deviation' and `detect the operation fraud'. Table~\ref{tab:scenario1} shows the instantiation of the proposed IoT-Pro context classification (Steps~5 and~6) in this scenario. 
Note that we assume that a single pick-up operation involves one truck, one driver and one cargo.
%For event context, we provide context identified and process events associated with them.

\begin{table}[t!!!]
\resizebox{\textwidth}{!}{%
\begin{tabular}{|c|l|l|l|l|l|}
\hline
\textbf{}                 & \multicolumn{1}{c|}{\textbf{Physical Objects}}                                                                            & \multicolumn{1}{c|}{\textbf{Location}}                                                              & \multicolumn{1}{c|}{\textbf{Time}}                                                                      & \multicolumn{1}{c|}{\textbf{Identity}}                                                   & \multicolumn{1}{c|}{\textbf{Environment}}                                                     \\ \hline
\multirow{4}{*}{Instance} & \begin{tabular}[c]{@{}l@{}}The truck is on the \\ blacklist or not\end{tabular}                                           & \multirow{4}{*}{Truck route at the port}                                                            & \begin{tabular}[c]{@{}l@{}}Total pickup \\ operation duration\end{tabular}                              & \multirow{4}{*}{Driver's credit at port}                                                 & \multirow{4}{*}{\begin{tabular}[c]{@{}l@{}}Weather \\ (has rain or not)\end{tabular}}         \\ \cline{2-2} \cline{4-4}
                          & \begin{tabular}[c]{@{}l@{}}The truck was \\ retrofitted or not\end{tabular}                                               &                                                                                                     & \begin{tabular}[c]{@{}l@{}}Empty truck \\ weighing duration\end{tabular}                                &                                                                                          &                                                                                               \\ \cline{2-2} \cline{4-4}
                          & Truck Category                                                                                                            &                                                                                                     & \begin{tabular}[c]{@{}l@{}}Loaded truck \\ weighing duration\end{tabular}                               &                                                                                          &                                                                                               \\ \cline{2-2} \cline{4-4}
                          & Cargo Weight                                                                                                              &                                                                                                     & Cargo loading duration                                                                                  &                                                                                          &                                                                                               \\ \hline
\multirow{8}{*}{Event}    & \begin{tabular}[c]{@{}l@{}}License plate number\\ - `evaluate the truck entry'\\ - `evaluate the truck exit'\end{tabular} & \begin{tabular}[c]{@{}l@{}}Truck entry location\\ - `evaluate the truck entry'\end{tabular}         & \begin{tabular}[c]{@{}l@{}}Truck arrival time\\ - start event\\ (the arrival of the truck)\end{tabular} & \begin{tabular}[c]{@{}l@{}}Driver's name\\ - `evaluate the truck entry'\end{tabular}     & \begin{tabular}[c]{@{}l@{}}Rain level\\ - `discontinue the pick-up \\ operation'\end{tabular} \\
                          & \begin{tabular}[c]{@{}l@{}}Empty truck weight\\ - `weigh the empty truck'\end{tabular}                                    & \begin{tabular}[c]{@{}l@{}}Empty truck weighing location\\ - `weigh the empty truck'\end{tabular}    & \begin{tabular}[c]{@{}l@{}}Empty truck weighing start time\\ - `weigh the empty truck'\end{tabular}     & \begin{tabular}[c]{@{}l@{}}Driver's ID\\ - `evaluate the truck entry'\end{tabular}       &                                                                                               \\
                          & \begin{tabular}[c]{@{}l@{}}Loaded truck weight\\ - `weigh the loaded truck'\end{tabular}                                  & \begin{tabular}[c]{@{}l@{}}Truck location when loading cargo\\ - `load in truck'\end{tabular}       & \begin{tabular}[c]{@{}l@{}}Empty truck weighing end time\\ - `weigh the empty truck'\end{tabular}       &                                                                                          &                                                                                               \\
                          &                                                                                                                           & \begin{tabular}[c]{@{}l@{}}Cargo location\\ - `load in truck'\end{tabular}                          & \begin{tabular}[c]{@{}l@{}}Cargo loading start time\\ - `load in truck'\end{tabular}                    &                                                                                          &                                                                                               \\
                          &                                                                                                                           & \begin{tabular}[c]{@{}l@{}}Loaded truck weighing location\\ - `weigh the loaded truck'\end{tabular} & \begin{tabular}[c]{@{}l@{}}Cargo loading end time\\ - `load in truck'\end{tabular}                      &                                                                                          &                                                                                               \\
                          &                                                                                                                           & \begin{tabular}[c]{@{}l@{}}Truck exit location\\ - `evaluate the truck exit'\end{tabular}           & \begin{tabular}[c]{@{}l@{}}Loaded truck weighing start time\\ - `weigh the loaded truck'\end{tabular}   &                                                                                          &                                                                                               \\
                          &                                                                                                                           &                                                                                                     & \begin{tabular}[c]{@{}l@{}}Loaded truck weighing end time\\ - `weigh the loaded truck'\end{tabular}     &                                                                                          &                                                                                               \\
                          &                                                                                                                           &                                                                                                     & \begin{tabular}[c]{@{}l@{}}Truck exit time\\ - end event\\ (truck leaves the port)\end{tabular}         &                                                                                          &                                                                                               \\ \hline
\multirow{2}{*}{Sensor}   & \begin{tabular}[c]{@{}l@{}}RFID\\ (attached to the truck)\end{tabular}                                                    & \multirow{2}{*}{GPS (signal)}                                                                       & \multirow{2}{*}{Timer}                                                                                  & \multirow{2}{*}{\begin{tabular}[c]{@{}l@{}}RFID \\ (attached to the truck)\end{tabular}} & \multirow{2}{*}{Rain sensor data}                                                             \\ \cline{2-2}
                          & Weigh sensors data                                                                                                        &                                                                                                     &                                                                                                         &                                                                                          &                                                                                               \\ \hline
\end{tabular}%
}
\vspace*{-.05\baselineskip}
\caption{Instantiation of the IoT-Pro context classification for Scenario 1}
\label{tab:scenario1}
\vspace*{-2\baselineskip}
\end{table}

% Based on the contexts instantiated in the IoT-Pro context classification, we select the following attributes to be added in the existing event log. We assume that a single pick-up operation involves one truck, one driver and one cargo. 
% Thus, \textit{the truck license plate number, driver ID, driver's credit in port (0: Low; 1: Medium; 2: High), truck in blacklist or not, truck getting retrofitted or not, truck's category (0-3: top 4 of most common trucks registered; 4: others), cargo location, weather (0: rain; 1: no rain) and cargo weight} will be introduced as case attributes in the existing event log. For event attributes captured by IoT data, \textit{truck\_location and truck\_weight} will be included.

Step~7 focuses on enriching the event log with new attributes to carry IoT-Pro context. 
The original event log has the following attributes: case id, activity, timestamp, and other existing non-IoT-related attributes provided in~\cite{Song2020FraudDO}. It is extended with the new attributes informed by the instantiation of the IoT-Pro context classification (see Table~\ref{tab:scenario1}), which then leads to an IoT-enriched event log for Scenario~1. Listing~\ref{lst:scenario_1} contains a conceptual Extensible Event Stream (XES)~\cite{xes_standard} schema of this IoT-enriched event log.

%In addition to the existing attributes in the event log (i.e., case id, activity, timestamp, and process-related attributes provided in~\cite{Song2020FraudDO}), it includes attributes extracted from the instantiation of IoT-Pro context classification (Step~7).

%In the following, we provide a conceptual Extensible Event Stream (XES)~\cite{xes_standard} schema (as shown in Listing~\ref{lst:scenario_1}) for an IoT-enriched event log generated in Scenario 1. In addition to existing attributes in the event log (i.e., case id, activity, timestamp, and process-related attributes provided in~\cite{Song2020FraudDO}), it includes attributes extracted from the instantiation of IoT-Pro context classification (Step~7).  In this use case, we assume that a single pick-up operation involves one truck, one driver and one cargo. 

\lstset{
    caption={Conceptual XES schema for the IoT-enriched event log in Scenario~1}, % Caption above the listing
    basicstyle=\ttfamily\scriptsize,
    breaklines=true,
    label={lst:scenario_1}, % Label for referencing this listing
    language=xes, % Use SQL functions/syntax highlighting
    frame=none, % Frame around the code listing
    showspaces=false,
    keywordstyle=\color{blue},
    commentstyle=\color{dkgreen},
    stringstyle=\color{mauve},
    showstringspaces=false,
    % Don't put marks in string spaces
    % multicols=2
}
\begin{lstlisting} 

<trace> 
<!-- Trace  attributes --> 
  <int key="case_id" value=? /> 
  <boolean key="customs_supervison" value=? /> /* 0: unsupervised; 1: supervised */
  <string key="cargo_type" value=? /> /* 0-3: top 4 of largest amount of cargo processed in port; 4: others */
  <float key="cargo_price" value=? /> 
  <string key="yard_category" value=? /> /* 0: outdoor yard; 1: warehouse; 2: others */
  <string key="means_of_payment" value=? /> /* 0: after operation completion; 1: monthly payment; 2: payment in advance */
  <string key="contract_category" value=? /> /* 0: for a single vessel; 1: long-term; 2: others */ 

  <!-- case attributes extracted from the IoT-Pro context classification -->
  <string key="truck_license_plate_number" value=? />
  <int key="driver_ID" value=? />
  <string key="driver_credit_in_port" value=? /> /* 0: Low; 1: Medium; 2: High */
  <boolean key="truck_blacklist" value=? /> /* 1: ever; 0: never */
  <boolean key="truck_retrofitted" value=? /> /* 0: retrofitted; 1: not retrofitted */
  <string key="truck_category" value=? /> /*  0-3: top 4 most common truck registered; 4: others */
  <string key="cargo_location" value=? />
  <string key="cargo_weight" value=? />
  <boolean key="weather" value=? /> /* 0: rain; 1:no rain */

  <event> 
    <!-- Event attributes --> 
    <date key="timestamp" value=? /> 
    <string key="activity"  value=? /> 
    
    <!-- Event attributes extracted from the IoT-Pro context classification -->
    <string key="truck_location" value=? /> 
    <string key="truck_weight" value=? /> 
  </event> 
  ...
</trace>  


\end{lstlisting}

% <int key = "cargo_temperature"  value = ? /> /* Value range: 0: <= 35 Celsius degree; 1: > 35 Celsius degree */
%     <string key = "cargo_humidity"  value= ? /> /* Value range: "not excessive"; "excessive" */
%     <string key = "cargo_smoke"  value = ? /> /* Value range: true; false */
%     <int key = "truck_temperature"  value = ? /> /* Value range: 0: <= 35 Celsius degree; 1: > 35 Celsius degree */

%\vspace*{-\baselineskip}
\paragraph{Discussion} 
%Since the timestamp is a mandatory attribute in the event log, the 
%(i.e., each event has a timestamp), we will not list this event attribute separately. In addition, 
Time duration can be calculated from timestamps stored in event logs. 
In order to simplify the complexity of the generated event log, the various duration will not be included as case attributes in the enriched event log. By comparing the truck path to a preset truck route based on domain knowledge, we may assess whether there are operational deviations in the process, such as whether the truck emptied the filler at a port location. However, the GPS data is so fine-grained that further study is required to understand how to process it. 
Hence, in this analysis scenario, we do not consider including the truck route as a case attribute in the event log.

% To evaluate the proposed IoT-Pro context classification, log-based queries are generated to check whether the generated IoT-enriched event logs contain sufficient data to answer the analysis questions. To test the query, we manually synthesise the event log. The query provided for examining the scenario~1 is shown in Figure~\ref{fig:query_s1}. It can also demonstrate that the generated IoT-enriched event logs can be used to analyse which factors are most likely to lead to the theft of cargo.

%\vspace*{-\baselineskip}
\subsubsection{Scenario~2}

During truck pick-up, operations may be suspended by the port company or the customer due to unsatisfactory or unsafe conditions for truck pick-up. The high temperatures in the coal yard or in the pick-up truck could be dangerous, and the high humidity of the coal is not cost-effective for the customer~\cite{SongCVHW22}. Therefore, a business concern (Step~1) in this case could be how to reduce the occurrence of such aborted operations to improve the efficiency of the truck pick-up process. An analytical question that corresponds to this business concern could be: \textit{Of the operations that come in for pick-up between 10 pm and 6 am, how many pick-up operations have been interrupted?}

% \begin{table}[b!!!]
% \vspace*{-\baselineskip}

% \usepackage{multirow}
% \usepackage{graphicx}
\begin{table}[]
\resizebox{\textwidth}{!}{%
\begin{tabular}{|c|l|l|l|}
\hline
\textbf{}                 & \multicolumn{1}{c|}{\textbf{Physical Object}}                                                                             & \multicolumn{1}{c|}{\textbf{Identity}}                                                  & \multicolumn{1}{c|}{\textbf{Environment}}                                                                                                                                                     \\ \hline
\multirow{8}{*}{Instance} & \multirow{8}{*}{\begin{tabular}[c]{@{}l@{}}The truck is on the \\ blacklist or not\end{tabular}}                          & \multirow{8}{*}{Driver's credit at port}                                                & Cargo condition                                                                                                                                                                               \\
                          &                                                                                                                           &                                                                                         & --- \textless{}=35 or \textgreater 35 Celsius degree                                                                                                                                          \\
                          &                                                                                                                           &                                                                                         & --- Excessive humid or not                                                                                                                                                                    \\
                          &                                                                                                                           &                                                                                         & --- Has smoke or not                                                                                                                                                                          \\
                          &                                                                                                                           &                                                                                         & Truck condition                                                                                                                                                                               \\
                          &                                                                                                                           &                                                                                         & --- \textless{}=35 or \textgreater 35 Celsius degree                                                                                                                                          \\
                          &                                                                                                                           &                                                                                         & Weather                                                                                                                                                                                       \\
                          &                                                                                                                           &                                                                                         & --- has rain or not                                                                                                                                                                           \\ \hline
\multirow{2}{*}{Event}    & \begin{tabular}[c]{@{}l@{}}License plate number\\ - `evaluate the truck entry'\\ - `evaluate the truck exit'\end{tabular} & \begin{tabular}[c]{@{}l@{}}Driver's name\\ - `evaluate the truck entry'\end{tabular}    & \begin{tabular}[c]{@{}l@{}}Temperature of the truck\\ Temperature of the cargo\\ Humidity of the cargo\\ Smoke of the cargo\\ Rain level\\ - `discontinue the pick-up operation'\end{tabular} \\
                          &                                                                                                                           & \begin{tabular}[c]{@{}l@{}}Driver's ID\\ - `evaluate the truck entry'\end{tabular}      &                                                                                                                                                                                               \\ \hline
\multirow{4}{*}{Sensor}   & \multirow{4}{*}{\begin{tabular}[c]{@{}l@{}}RFID\\ (attached to the truck)\end{tabular}}                                   & \multirow{4}{*}{\begin{tabular}[c]{@{}l@{}}RFID\\ (attached to the truck)\end{tabular}} & Temperature sensor data                                                                                                                                                                       \\
                          &                                                                                                                           &                                                                                         & Humidity sensor data                                                                                                                                                                          \\
                          &                                                                                                                           &                                                                                         & Smoke sensor data                                                                                                                                                                             \\
                          &                                                                                                                           &                                                                                         & Rain sensor data                                                                                                                                                                              \\ \hline
\end{tabular}%
}
\vspace*{.05\baselineskip}
\caption{Instantiation of the IoT-Pro context classification for Scenario 2}
\vspace*{-2\baselineskip}
\label{tab:scenario2}
\end{table}

% \caption{Instantiation of the IoT-Pro context classification for Scenario 2}
% \vspace*{-2\baselineskip}
% \label{tab:scenario2}
% \end{table}

The suspension of pick-up operations depends mainly on the condition of the trucks, the condition of the cargo and the weather. This information can be captured by environmental sensors such as temperature, humidity, smoke and rain sensors. Another reason for the unexpected termination of the pick-up process is that credit problems with the truck or driver prevent them from being allowed to pick up the cargo. This could be indicated through the information stored in the database by reading RFID tags attached on the windshield of the truck (Step~3). In the IoT-enhanced truck pick-up process (see Fig.~\ref{fig:iot_enhanced_truck_bpmn}), an activity that uses this IoT data is `Discontinue the pick-up operation' (Step~4) . Table~\ref{tab:scenario2} illustrates the instantiation of the proposed IoT-Pro context classification (Steps~5 and~6) for this scenario. 
Based on this instantiation, the context to be carried by new case attributes to enrich the event log for this scenario contains:   \texttt{truck\_license\_plate\_number}, \texttt{driver\_ID}, \texttt{driver\_credit\_in\_port}, \texttt{truck\_blacklist}, \texttt{cargo\_temperature}, 
\texttt{cargo\_smoke}, \texttt{cargo\_humidity}, \texttt{truck\_temperature} and \texttt{weather}. 
%\vspace*{-2\baselineskip}

\begin{figure}[h!]
    \vspace*{-1\baselineskip}
    \centering
    \includegraphics[width=\textwidth]{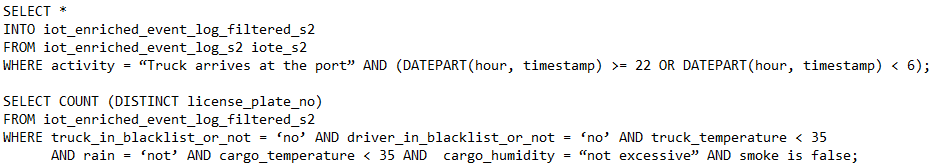}
    \vspace*{-1.5\baselineskip}
    \caption{log\_based query for Scenario 2}
\vspace*{-2\baselineskip}
\label{fig:query_s2}
\end{figure}

\newpage
To evaluate the proposed method, we have generated log-based queries (as depicted in Fig.~\ref{fig:query_s2}) for this scenario to check whether the yielded IoT-enriched event logs contain sufficient data to answer the analytical question. 
% For future works, we will synthesise an event log with the knowledge of domain experts to further examine the proposed method.

% \begin{figure}[h!]
%     \centering
%     \includegraphics[width=\textwidth]{figs/sql_query_scenario1.png}
%     \caption{log\_based query for Scenario 1}
%     \label{fig:query_s1}
% \end{figure}

	% 5 pages including figures	
\section{Conclusions and Future Work} 
\label{sec:conclusion}

In this work, we have proposed a method for deriving IoT-enriched event logs, named AMORETTO. The method consists of two phases. %In the first phase, 
First, we construct the IoT-Pro context classification using an established taxonomy development method. 
Second, we present a method for integrating IoT data with event logs, guided by IoT-Pro. As a result, this will yield an IoT-enriched event log. We have demonstrated the applicability of AMORETTO with a real-life use case. 
%In particular, we provide two analysis scenarios, each requiring the inclusion of specific IoT contexts into the event log in order to answer a specific analysis question. In addition, we have offered log-based queries to determine if the generated IoT-rich event logs contain sufficient data to answer certain analytical questions.
%sensor context, event context and instance context which have informed the creation of new attributes for enriching event logs with corresponding IoT data. 

%\vspace*{-.5\baselineskip}
%\paragraph{Limitations} 
However, our work has several limitations which can be addressed in the future work. 
First, when developing the IoT-Pro context classification, we mainly adopted the ``conceptual-to-empirical'' approach proposed in the existing taxonomy development method~\cite{Nickerson2013AMF}. We refer to our findings as a classification rather than a taxonomy because it has not been empirically examined. 
Therefore, to conduct the ``empirical-to-conceptual'' analysis with a good amount of real-world use cases and data is a direction for future work. 

Second, using the proposed IoT-Pro context classification, IoT data can conceptually be captured into all levels of process context. However, from the two analysis scenarios in Sect.~\ref{sec:scenarios}, only relevant event context and instance context can be included in event logs as new attributes to carry the corresponding IoT data. Hence, we will investigate how the process context and organisational context in the proposed IoT-Pro context classification can be incorporated into the event log.

Third, in this work, an event log follows an XES schema, and this schema is only capable of capturing intra-case data~\cite{SenderovichFGJM17}, such as event context and instance context. 
Process context provide an aggregated view of instance context and this is considered as inter-case data~\cite{SenderovichFGJM17}, which cannot be captured by an XES schema. 
To this end, we will investigate how to extend current XES schema to incorporate IoT-Pro context.

Last but not least, we plan to acquire and collect domain knowledge into synthesis of an event log to evaluate the performance of the proposed method. 
Furthermore, we will examine and evaluate the proposed method considering a specific application domain such as the healthcare domain.

%
%In addition, organisational context carries business rules related to processes. Whilst it cannot be derived from sensor context directly, organisational context can provide guidance for process execution and support derivation of context at other levels. 

%together with other levels of context in the proposed IoT-Pro context classification. Therefore, the inclusion of process context and organisational context into the current event log schema requires further research.

%In addition, the lowest level of context in the proposed IoT-Pro context classification is the sensor context. However, there is more fine-grained data related to sensors, such as a sensor's working state and location. This is out of the scope of this work but will be investigated in the future.

%\vspace*{-.5\baselineskip}
%\paragraph{Future work} 

%%% --------- TEMPORARILY COMMENTED OUT --------- %%% 
%We further demonstrate its applicability by applying AMORETTO to a real-life use case. In particular, we provide two analysis scenarios, each requiring the inclusion of specific IoT contexts into the event log in order to answer a specific analysis question. In addition, we have offered log-based queries to determine if the generated IoT-rich event logs contain sufficient data to answer certain analytical questions.

\vspace*{-.5\baselineskip}
\subsubsection{Acknowledgement:} 
We thank Rongjia Song, Weiping Cui, Jan Vanthienen, Lei Huang and Ying Wang for providing a detailed discussion about the truck pick-up cargo process at a bulk port in China in their recent publication~\cite{SongCVHW22}, which we have adopted as a real-life use case for demonstrating the applicability of AMORETTO in our work.

%We particularly thank Manuel Camargo, Marlon Dumas, and Oscar González Rojas for the high qualityode they released which allowed fast reproduction of the experimental setting and the processing of event logs.
 % 1 page
% -----sections/references. 1-2 pages

%
% ---- Bibliography ----		% p16-p17
\bibliographystyle{splncs04}
\bibliography{reference}
\end{document}